\documentclass[]{rQUF2e}
\bibpunct{(}{)}{,}{a}{}{,}

\newcommand{\eg}{{\it e.g.}~}

\newcommand{\ave}[1]{\left\langle#1 \right\rangle}

\newcommand{\elabel}[1]{\label{eq:#1}}
\newcommand{\eref}[1]{(Eq.~\ref{eq:#1})}

\newcommand{\Eref}[1]{Equation~(\ref{eq:#1})}

\newcommand{\flabel}[1]{\label{fig:#1}}
\newcommand{\fref}[1]{Fig.~\ref{fig:#1}}

\newcommand{\Ito}{It\^{o}}

\begin{document}
\doi{10.1080/1469768YYxxxxxxxx}
 \issn{1469-7696} \issnp{1469-7688} \jvol{00} \jnum{00} \jyear{2008} \jmonth{July}

\markboth{O. Peters}{Optimal leverage from non-ergodicity}
\title{Optimal leverage from non-ergodicity} 

\author{Ole Peters$^{\ast}$\thanks{$^\ast$Ole Peters. Email: ole@santafe.edu
\vspace{12pt}}\\  
\normalfont{
Department of Mathematics and Grantham Institute for Climate Change,
Imperial College London,
180 Queens Gate,
London SW7 2AZ,
United Kingdom\\
and\\
Department of Atmospheric Sciences,
University of California Los Angeles,
7127 Math. Sci. Bldg.,
405 Hilgard Ave.,
Los Angeles, CA, 90095-1565, USA,\\
and\\
Santa Fe Institute,
1399 Hyde Park Rd., 
Santa Fe, NM, 87501, 
USA
}
\vspace{12pt}\received{Received: 13 February 2009; in final form: 17 June 2010} }

\maketitle

\begin{abstract}
In modern portfolio theory, the balancing of expected returns on
investments against uncertainties in those returns is aided by the use
of utility functions. The Kelly criterion offers another approach,
rooted in information theory, that always implies logarithmic
utility. The two approaches seem incompatible, too loosely or too
tightly constraining investors' risk preferences, from their respective
perspectives. The conflict can be understood on the basis that the
multiplicative models used in both approaches are non-ergodic which
leads to ensemble-average returns differing from time-average returns
in single realizations. The classic treatments, from the very
beginning of probability theory, use ensemble-averages, whereas the
Kelly-result is obtained by considering time-averages. Maximizing the
time-average growth rates for an investment defines an optimal
leverage, whereas growth rates derived from ensemble-average returns
depend linearly on leverage. The latter measure can thus incentivize
investors to maximize leverage, which is detrimental to time-average
growth and overall market stability. The Sharpe ratio is insensitive
to leverage. Its relation to optimal leverage is discussed. A better
understanding of the significance of time-irreversibility and
non-ergodicity and the resulting bounds on leverage may help policy
makers in reshaping financial risk controls.
\end{abstract}

\newpage
\vspace{-50pt}

\section*{}
This study focuses on the simple setup of self-financing investments,
that is, investments whose gains and losses are reinvested without
consumption or deposits of fresh funds, in assets undergoing geometric
Brownian motion. The consequences of time irreversibility pertaining
to studies of risk are discussed. Understanding these consequences
appears particularly important in the light of the current financial
and economic crisis. This will be elaborated at the end, in
Sec.~\ref{Discussion}, after establishing the main concepts.

In Sec.~\ref{Introduction} the portfolio selection problem, as
introduced by \cite{Markowitz1952}, is reviewed. Its use of utility to
express risk preferences is contrasted with a different ansatz,
proposed by \cite{Kelly1956}, that makes use solely of the r\^ole of
time in multiplicative processes. While in the terminology of modern
portfolio theory, the latter ansatz can be interpreted as the
assumption of logarithmic utility, in Sec.~\ref{Two_averages} the
Kelly result is shown to be equivalent, in the present setup, to an
application of It\^{o}'s formula of stochastic calculus. In this sense
it is not the reflection of a particular investor's risk preferences
but a generic null-hypothesis. Considerations of personal risk
preferences can improve upon this hypothesis but they must not obscure
the crucial r\^ole of time. In Sec.~\ref{Ergodicity} it is shown by
explicit calculation that the non-ergodicity of geometric Brownian
motion can create a difference between ensemble-average and
time-average growth rates. \Ito's formula is seen as a means to
account for the effects of time. In Sec.~\ref{Time_scales} the
growth-optimal leverage, which specifies a portfolio along the
efficient frontier, is derived and related to a minimum investment
time-horizon. Optimal leverage is compared to the Sharpe
ratio. Finally, in Sec.~\ref{Discussion} implications of the results
from Sec.~\ref{Time_scales} for real investments are discussed, and
the concept of statistical market efficiency is introduced.

\section{Introduction}
\label{Introduction}
Modern portfolio theory deals with the allocation of funds among
investment assets. We assume zero transaction costs and portfolios
whose prices $p(t)$ follow geometric Brownian motion\footnote{Some
  authors define the parameters of geometric Brownian motion
  differently \citep{Timmermann1993}.  The parameters in their notation
  must be carefully translated for comparisons.},
\begin{equation}
dp(t)=p(t)\left(\mu dt+\sigma dW_t\right),
\elabel{motion}
\end{equation}
where $\mu$ is a drift term, $\sigma$ is the volatility, and
\begin{equation}
W(T)\equiv \int_{t=0}^{t=T}dW_t
\elabel{finite}
\end{equation}
is a Wiener process.

\cite{Markowitz1952} suggested to call a portfolio $i$
efficient if

\noindent a) there exists no other portfolio $j$ in the market with
equal or smaller volatility, $\sigma_j \leq \sigma_i$, whose drift
term $\mu_j$ exceeds that of portfolio $i$,
\begin{equation}
\text{for all~} j~ \text{such that~} \sigma_j\leq\sigma_i, \text{~we have~} \mu_j\leq \mu_i.
\elabel{efficient1}
\end{equation}
b) there exists no other portfolio $j$ in the market with equal or
greater drift term, $\mu_j \geq \mu_i$, whose volatility $\sigma_j$ is
smaller than that of portfolio $i$,
\begin{equation}
\text{for all~} j~ \text{such that~} \mu_j\geq \mu_i, \text{~we have~} \sigma_j\geq \sigma_i.
\elabel{efficient2}
\end{equation}
\cite{Markowitz1952} argued that it is unwise to invest in any
portfolio that is not efficient. In the presence of a riskless asset
(with $\sigma_i$=0) all efficient portfolios lie along a straight line
-- the efficient frontier -- that intersects, in the space of
volatility and drift terms, the riskless asset, $R$, and the
so-called market portfolio, $M$, \citep{Tobin1958}, see
\fref{surface}.

Markowitz' suggestion to focus on the mean $\mu$ and the variance
$\sigma^2$ was later criticized because the first two moments alone do
not sufficiently constrain the return distribution. Clearly preferable
portfolios can appear inferior if assessed only by Markowitz'
1952-criteria \citep{HanochLevy1969}. Below it will be stated
specifically in what sense large $\mu$ and small $\sigma$ are
desirable under the dynamics of \eref{motion}.

Since any point along the efficient frontier represents an efficient
portfolio, Markowitz' (1952) arguments need to be augmented with
additional information in order to select the optimal portfolio. This
additional information is generally considered a property of the
investor, namely his risk preference, represented by a utility
function, $u=u\left(p(t)\right)$, that specifies the usefulness or
desirability of a particular investment outcome to a particular
investor\footnote{The concept of assigning a utility to a payoff from
  uncertain investments can be traced back to Bernoulli's
  St. Petersburg Paradox \citep{Bernoulli1738}. The paradox captures
  the essence of the problem treated here: is an investment with
  infinite expected pay-off worth an infinite risk?  The recognition
  of non-ergodicity, as will be shown elsewhere, also resolves this
  paradox.}.

\begin{figure}
\flabel{surface}
\begin{center}
\includegraphics[scale=0.5,angle=0]{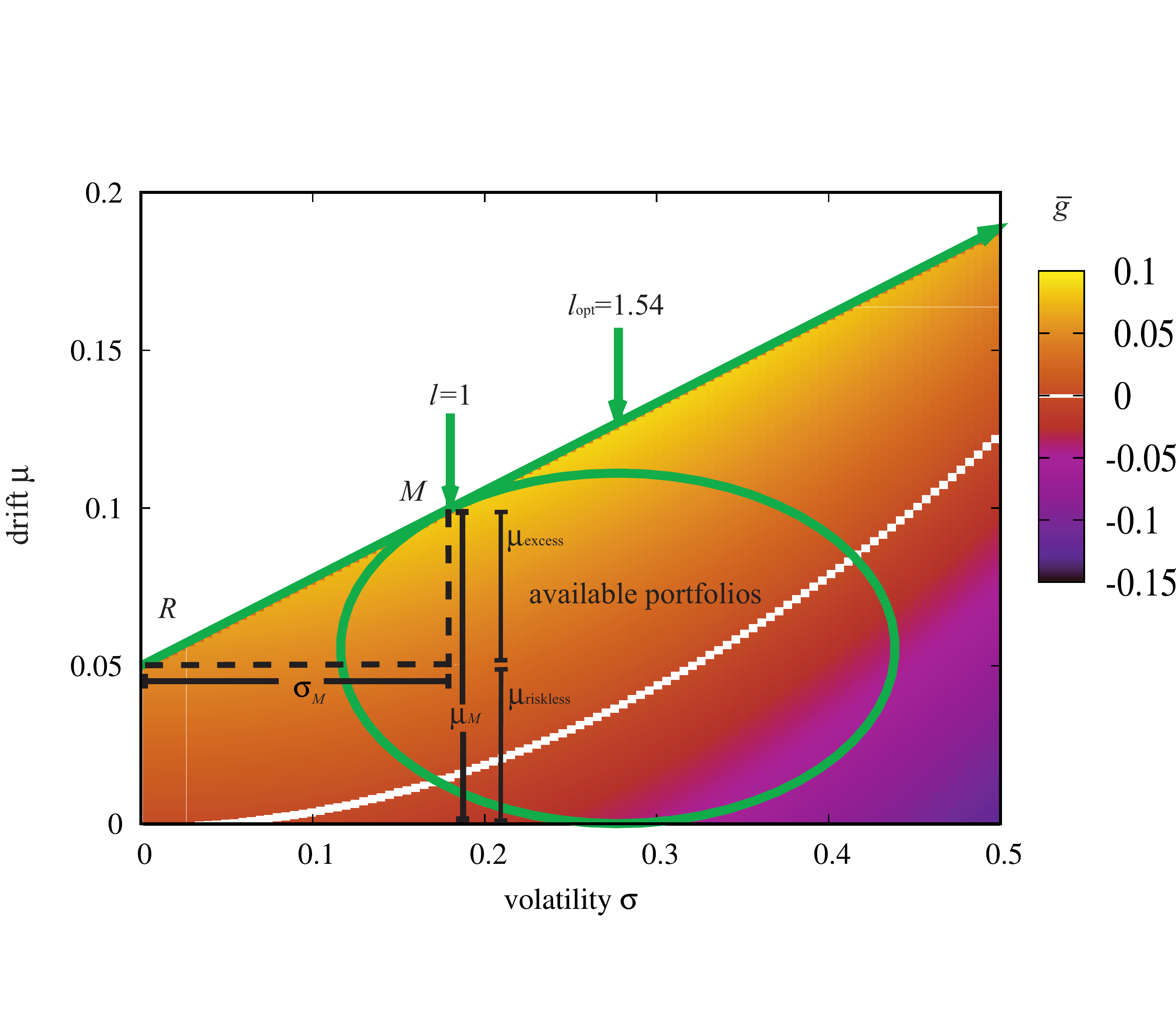}
\end{center}
\caption{The efficient frontier (green straight line) intersects the
  the riskless asset $R$ (here $\mu_{\text{riskless}}=0.05$ per time
  unit) and is tangent to the space of available portfolios (green
  oval), touching at the point defined as the market portfolio, $M$
  (here $\mu_{M}=0.1$ per time unit, resulting in an excess expected
  return of $M$ compared to $R$ of $\mu_{\text{excess}}=0.05$ per time
  unit; the volatility of the market portfolio $\sigma_M=0.18$ per
  square-root of time unit), corresponding to leverage $l=1$, left
  arrow. The color coding shows the expected, more specifically
  time-average, growth rate, $\bar{g}=\mu-\sigma^2/2$; the portfolio
  of optimal leverage (see Sec.~\ref{Time_scales}, here
  $l_{\text{opt}}\approx 1.54$) along the efficient frontier is
  indicated by the right arrow. Both for fixed volatility $\sigma$ and
  fixed expected return, which we call the ensemble-average growth
  rate, $\mu$, there are no obtainable portfolios (those below the
  efficient frontier) whose time-average growth rates exceed that at
  the efficient frontier. Zero time-average growth rate is indicated
  white.}
\end{figure}

In a parallel development, \cite{Kelly1956} considered portfolios that
were also described by two parameters. In his case, the portfolios
were double-or-nothing games on which one could bet an arbitrary
fraction of one's wealth (one parameter) and knew the outcome with
some probability (second parameter).
Both \cite{Markowitz1952} and \cite{Kelly1956} recognized that it is
unwise to maximize what is often called the expected rate of return,
\begin{equation}
\ave{g}=\frac{1}{dt}\ave{\frac{dp(t)}{p(t)}},
\elabel{return}
\end{equation}
where $\ave{}$ denotes the ensemble mean over realizations of the
Wiener process. \cite{Markowitz1952} rejected such strategies because
the portfolio with maximum expected rate of return is likely to be
under-diversified.
In Kelly's case the probability of bankruptcy approaches one as games
of maximum rate of return are repeated \citep{Kelly1956}. In geometric
Brownian motion bankruptcy is impossible, but the effects of time are
essentially the same as in Kelly's setup.

While Markowitz emphasized parameters such as risk preferences and
personal circumstances (``The proper choice among portfolios depends
on the willingness and ability of the investor to assume risk.''
\citep{Markowitz1991}), Kelly used a fundamentally different ansatz by
maximizing the so-called expected growth rate,
\begin{equation}
\bar{g}=\frac{1}{dt}\ave{d \ln p},
\elabel{Kelly}
\end{equation}
rather than the expected rate of return, without an {\it a priori}
need for additional information. The exact meaning of these two
quantities will be worked out in Sec.~\ref{Ergodicity}, and a more
precise nomenclature will be introduced shortly. The conditions under
which the growth-rate ansatz alone yields meaningful results have been
discussed in the literature
\citep{MertonSamuelson1974,Markowitz1976,Markowitz1991}. For
self-financing portfolios (the focus of this study), where eventual
outcomes are the product over intermediate returns, these conditions
are met. This is a good approximation, \eg for large pension funds
where fluctuations in assets under management are dominated by market
fluctuations \citep{SchwarzkopfFarmer2008} and, arguably, for entire
economies. Some stock market indeces, for example the DAX, also
reflect the value of a hypothetical constant rebalanced self-financing
portfolio with zero transaction costs. \Eref{Kelly} shows that the
Kelly criterion, maximizing the expected growth rate, is
mathematically similar to using logarithmic utility. In this special
case, {\it i.e.}  $u(p(t))=\ln\left(p(t)\right)$, the ensemble-average
of the utility happens to be the time-average of the growth rate in a
multiplicative process. The fact that the process is non-ergodic and
the time-average has to be used explains why logarithmic utility so
often yields intuitively sensible results, see Sec.~\ref{Ergodicity}.

The so-called Sharpe ratio, usually defined as
$S=(\ave{g}-\mu_{\text{riskless}})/\sigma$, where $\mu_{\text{riskless}}$
is the rate of return on a riskless asset, is a means of analysis
using Markowitz' framework. It can be thought of as the slope of a
straight line in \fref{surface} intersecting the riskless asset. We
will return to the Sharpe ratio in Sec.~\ref{Sharpe_Ratio}, as it is
best discussed using the main results about to be presented.

\subsection{Two averages}
\label{Two_averages}
In this section the reader is reminded that the two averages
  \eref{return} and \eref{Kelly} are not necessarily identical. For
riskless assets, the chain rule of ordinary calculus implies that
\eref{return} and \eref{Kelly} are identical,
$\ave{g}_{\text{riskless}}=\bar{g}_{\text{riskless}}$, but this
  is not the case for non-zero volatility.

Combining \eref{motion} and \eref{return}, we now compute the
expectation value of the fractional price increment per infinitessimal
time step, the expected rate of return,
\begin{align}
\ave{g}
&=\frac{1}{dt}\ave{\frac{p(t)\mu dt+p(t)\sigma dW_t}{ p(t)}}\elabel{expected} \\
&= \mu +\sigma  \frac{1}{dt}\ave{dW_t} \nonumber \\
&=\mu.  \nonumber
\end{align}
From now on we will call this quantity the ensemble-average growth
rate, for reasons that will be made clear in Sec.~\ref{Ergodicity}.

The object $d \ln p$ in \eref{Kelly} has to be treated carefully using
It\^{o}'s formula\footnote{We stress that It\^{o}'s interpretation of
  increments like \eref{motion} is indeed the appropriate choice in
  the present context because it implies statistical independence of
  $p(t)$ and the increment $dW_t$ and no knowledge of the
    future. Alternative interpretations are possible, notably
  Stratonovich's, but they define different dynamics. For a detailed
  discussion, see \cite{vanKampen1992}, Ch.~9, \cite{LauLubensky2007}
  and ~\cite{Oksendal2005}, Ch.~3 and Ch.~5.}. With the chain rule of
ordinary calculus replaced by It\^{o}'s version, \eref{return} and
\eref{Kelly} now correspond to different averages.

It\^{o}'s formula for \eref{motion}, which we need to evaluate
\eref{Kelly}, takes the form\footnote{This calculation can be found in
any textbook on financial derivatives, \eg \cite{Hull2006}
Ch.~12.}
\begin{equation}
df=\left(\frac{\partial f}{\partial t}+\mu p\frac{\partial f}{\partial p}+\frac{1}{2}\sigma^2 p^2\frac{\partial^2 f}{\partial p^2} \right) dt+p\sigma \frac{\partial f}{\partial p} dW_t,
\elabel{formula}
\end{equation}
where $f=f(p(t),t)$ is some function of the It\^{o} process $p(t)$ of
\eref{motion}, and time $t$. The dependencies of $p(t)$ and
$f(p(t),t)$ have been left out in \eref{formula} to avoid clutter. The
third term on the right of \eref{formula} constitutes the difference
from the increment for a function of a deterministic process. Due to
the second derivative, It\^{o}'s formula can only take effect if
$f(p(t),t)$ is non-linear in $p$. To derive the increment $d \ln p$,
we need to choose $f(p(t),t) \equiv \ln (p(t))$, that is, a non-linear
function. We arrive at
\begin{equation}
d\ln p=\left(\mu-\frac{\sigma^2}{2}\right)dt+\sigma dW_t.
\elabel{log_motion}
\end{equation}
Notice that \Ito's formula changes the behavior in time, whereas the
noise term is unchanged. In the literature, the corresponding average,
\begin{align}
\bar{g}&=\frac{1}{dt}\ave{d\ln p}\elabel{growth_rate}\\
&=\mu-\frac{\sigma^2}{2}+\frac{\sigma}{dt} \ave{dW_t} \nonumber\\
&=\mu-\frac{\sigma^2}{2},\nonumber
\end{align}
is called the expected growth rate, or logarithmic geometric mean rate
of return. Here we call it the time-average growth rate.

Distributions of logarithmic returns for many asset classes are highly
non-Gaussian, see \eg \cite{MantegnaStanley1995}. This does not affect
the applicability of the concepts about to be discussed, however, as
their justification is the irreversibility of time, see
Sec.~\ref{Ergodicity}. In general, the time-average growth rate of a
self-financed portfolio whose rates of return obey a given probability
distribution is the logarithm of the geometric mean of that
distribution, see \eg \cite{Kelly1956,Markowitz1976}. Extending the
results of this study to return distributions that are not log-normal
thus only requires the computation of the geometric mean.

\Eref{growth_rate} shows that while the ensemble-average growth rate
enters into the time-average growth rate, it does so in combination
with the volatility, quantifying for the
present setup the statement that large returns and small volatilities
are desirable. This is illustrated in \fref{surface}. 

\section{Ergodicity}
\label{Ergodicity}
How can we make sense of the difference between the quantities
computed in \eref{expected} and \eref{growth_rate} in the presence of
non-zero volatility? The problem that additional information is needed
to select the right portfolio, which was first treated by Bernoulli
(1738), disappears when using \eref{growth_rate} -- how did this
problem arise in the first place, and what is the meaning of
\eref{expected}? It will be shown in this section that the non-ergodic
nature of \eref{motion} allows us to obtain $\ave{g}=\mu$ from an
estimate for the growth rate averaged over an ensemble of infinitely
many realizations of the stochastic process, whereas the time average
of the same estimate produces $\bar{g}=\mu-\frac{\sigma^2}{2}$.

To generate the ensemble average, stochasticity is removed by letting
the sample size, or number of realizations, or number of parallel
universes diverge before the non-trivial effects of time, which arise
from multiplicative noise, are fully taken into account.
\Eref{expected} is thus the answer to the following question: ``what
is the rate of return on this investment, computed from an average
over all possible universes?'', where a universe is defined as a
particular sequence of events, {\it i.e.} one realization of the
process \eref{motion}. This nomenclature is employed to emphasize that
we only live in one realization of the universe but stay alive in that
universe for some time. For most of us, therefore, this question is
less relevant than the question: ``what is the rate of return on this
investment, averaged over time?'', to which the answer is
\eref{growth_rate}. This is illustrated in \fref{universes}, where one
realization of a self-financed portfolio is compared to an average
over an increasing number of universes. This is achieved by producing
independent sequences of wealth, corresponding to resetting an
investor's wealth and starting over again. The independent sequences
are then averaged (arithmetically) at equal times. As this averaging
procedure over universes destroys stochasticity, the stochastic
exponential growth process (whose time-average growth rate is
\eref{growth_rate}) approaches deterministic exponential growth (whose
growth rate is the ensemble-average growth rate, \eref{expected}). The
procedure of starting over again is like going back in time, or
periodically resetting one's investment to its initial value. But
going back in time is not possible, and the self-financing portfolios
considered here do not allow any
resetting\footnote{\label{horse_note}There are situations in which the
  ensemble average is more relevant. Kelly constructed the example of
  a gambler whose wife, once a week, gives him an allowance of one
  dollar to bet on horses \citep{Kelly1956}. The optimal strategy for
  this gambler is to maximize the expected return,
  \eref{expected}. The reason is that the gambler resets his wealth in
  each round of the game instead of reinvesting. His wealth is the sum
  (a linear object) of past gains, whereas under re-investment it
  would be the product (an object that is non-linear and hence
  affected by It\^{o}'s formula). In Sec.~\ref{Time_scales} we will
  see how this translates into preferred values of leverage.}. For
processes with Wiener noise, \eref{finite}, It\^{o}'s formula can
encode the multiplicative effect of time in the ensemble
average. These intuitive arguments will now be made precise.

\begin{figure}
\flabel{universes}
\begin{center}
\includegraphics[scale=0.5,angle=0]{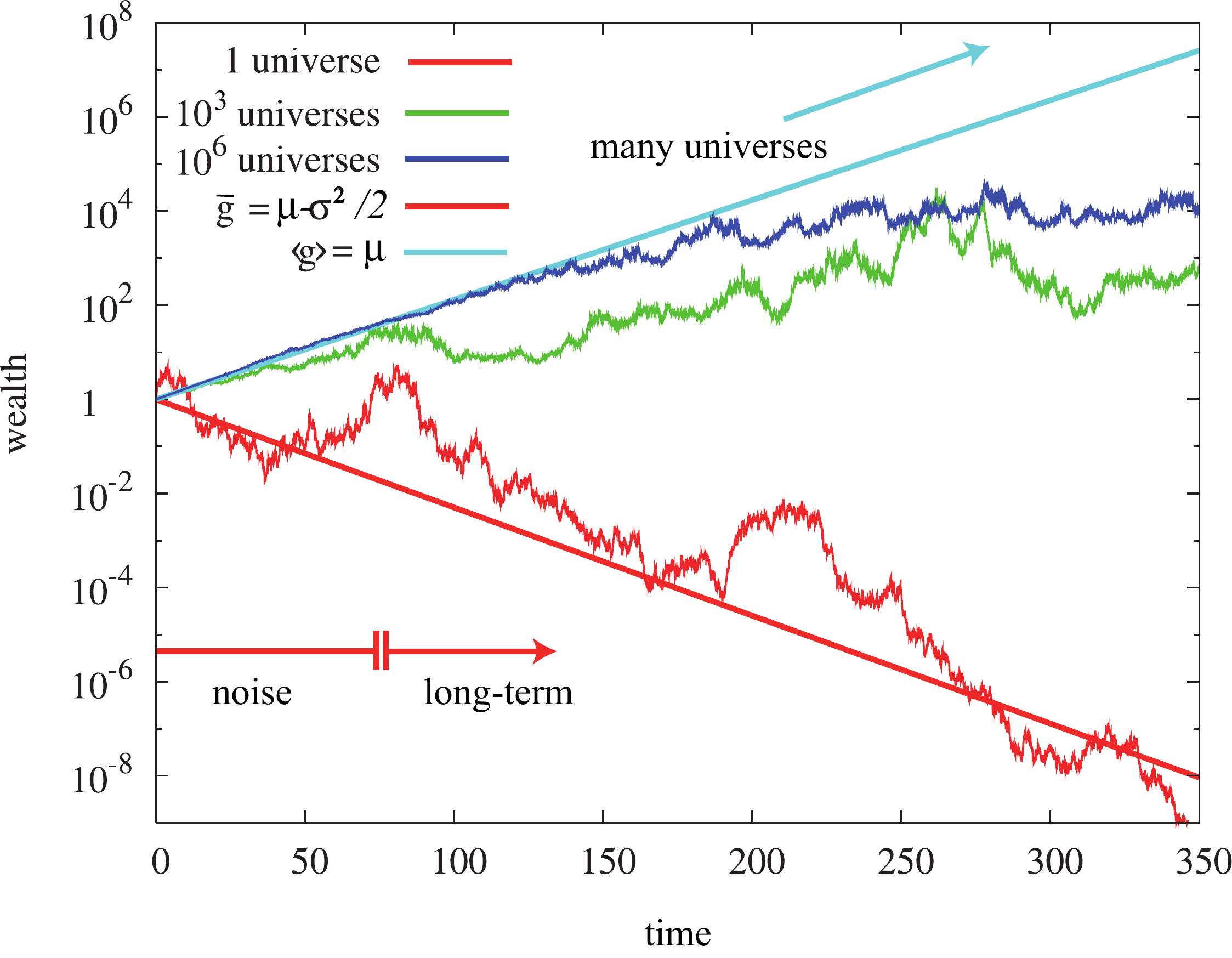}
\end{center}
\caption{Wealth starts at unity at time $T=0$ and then behaves
  according to \eref{motion} with $\mu=0.05$ per time unit and
  volatility $\sigma=0.45$ per square root of one time unit, implying
  the time-average exponential growth rate $\bar{g}\approx -0.051$ per
  time unit. For short times the performance is noise-dominated; after
  $T\approx 75$ time units (red arrow, \eref{long_term}) the
  time-average growth rate takes over (see also
  \fref{relative_error}). To uncover the ensemble-average growth rate
  \eref{return} from the dynamics, an average over many independent
  universes must be taken.}
\end{figure}

Ergodicity requires a unique stationary probability distribution of
the process $p(T)$ in the long-time limit $T \to \infty$. Geometric
Brownian motion is therefore trivially non-ergodic because it is not a
stationary process. This implies that there is no guarantee for the
ensemble average of an observable to be identical to its time average.
The observable we are interested in is the growth rate. It will be
shown that \eref{return} corresponds to the ensemble average of a
particular estimator for the growth rate, and that \eref{Kelly}
corresponds to the time average.

In practice, an exponential growth rate is estimated from observations
over a finite time $T$. To add the possibility of averaging over $N$
parallel universes (or completely independent systems), we consider
the estimator
\begin{equation}
g_{\text{est}}(T,N)=\frac{1}{T} \ln \ave{\frac{p(T)}{p(0)}}_N.
\elabel{growth_estimate}
\end{equation}
Here, the angled brackets denote the average over $N$ realizations,
$\ave{}_N=\frac{1}{N}\sum_{i=1}^N$. The different $p_i(T)$ are
obtained by solving the stochastic differential equation \eref{motion}
by integrating \eref{log_motion} over time and then exponentiating,
\begin{equation}
p_i(T)=p(0) \exp\left(\left(\mu-\frac{\sigma^2}{2}\right)T+\sigma W_i(T)\right).
\elabel{solution}
\end{equation}
The sample average in \eref{growth_estimate} must not enclose the
logarithm. The quantity that reveals the non-ergodic properties of
\eref{motion} and clarifies the meaning of \eref{return} is obtained
by averaging the values $p_i(T)$ from individual realizations, $i$,
first, before the logarithm translates them into a growth rate. This
procedure corresponds precisely to \fref{universes}, where outcomes
$p_i(T)$ are averaged first at equal times, and then a growth rate is
derived by taking the logarithm. This is different from Bernoulli's
treatment, where the logarithm is a utility function and would be
inside the sample average, obscuring the conceptual failure of the
ensemble-average. It was \cite{Kelly1956} who first pointed out that
the time average should be considered instead.

The ensemble average of the estimator \eref{growth_estimate} can be
identified with the limit
\begin{align}
\ave{g}=\lim_{N\to \infty} g_{\text{est}}(T,N),
\elabel{ensemble_average}
\end{align}
whereas the time average results from the limit
\begin{equation}
\bar{g}=\lim_{T\to\infty} g_{\text{est}}(T,N=1).
\elabel{time_average}
\end{equation}
Writing the averages as these two limits helps elucidate the relation
between \eref{return} and \eref{Kelly}, and it shows the symmetry or
absence thereof between effects of additional time and effects of
additional parellel universes included in the estimate. We will now
calculate both to show explicitly that the limits are not
interchangeable.
We start with the ensemble-average. The Wiener process $W(T)$ in
\eref{finite} is Gaussian-distributed with mean 0 and standard
deviation $\sqrt{T}$. Using the fundamental transformation law of
probabilities, \eref{solution} thus implies that
$\left(\frac{p(T)}{p(0)}\right)$ is log-normally distributed, according to
\begin{equation}
P\left(\frac{p(T)}{p(0)}\right)=\frac{1}{\left(\frac{p(T)}{p(0)}\right)\sqrt{2\pi T \sigma^2}}\exp\left(-\frac{\left(\ln\left(\frac{p(T)}{p(0)}\right)-(\mu-\frac{\sigma^2}{2})T\right)^2}{2 T \sigma^2}\right).
\end{equation}
The first moment of
$\left(\frac{p(T)}{p(0)}\right)$ is
\begin{equation}
\lim_{N\to\infty}\ave{\frac{p(T)}{p(0)}}_N=\exp(\mu T),
\end{equation}
the well known expectation value of log-normally distributed
variables. Using this in \eref{ensemble_average} in conjunction with
\eref{growth_estimate} yields
\begin{align}
\ave{g}&=\mu \elabel{ensemble_average3}\\
&=\frac{1}{dt}\ave{\frac{dp(t)}{p(t)}},\nonumber
\end{align}
where the last line follows from \eref{expected}, explaining our
nomenclature of referring to \eref{return} as an ensemble-average
growth rate.

Next, we consider \eref{time_average}, where the long-time limit is
responsible for eliminating stochasticity. Using $N=1$ since we are
interested only in one realization that could be our reality, and
substituting \eref{growth_estimate} and \eref{solution} in
\eref{time_average} we find
\begin{align}
\bar{g}&=\lim_{T\to\infty}\frac{1}{T}\left(\left(\mu-\frac{\sigma^2}{2}\right)T+\sigma W(T)\right)\elabel{time_average2}\\
&=\mu-\frac{\sigma^2}{2}+\lim_{T\to\infty}\left(\sigma T^{-1/2}W(1)\right)\nonumber\\
&=\mu-\frac{\sigma^2}{2}\nonumber\\
&=\frac{1}{dt}\ave{d \ln p}.\nonumber
\end{align}
The step from line one to two in \eref{time_average2} follows from the
scaling properties of Brownian motion. Although clearly $W(T)$ cannot
be equated to $T^{1/2}W(1)$ for any specific realization, the step is
valid because of the limit $T\to\infty$. The final line follows from
\eref{growth_rate}, and we identify the time-average growth rate with
\eref{Kelly}. The decay of the stochastic term as $T^{-1/2}$ is
illustrated in \fref{relative_error}.

\begin{figure}
\begin{center}
\includegraphics[scale=0.5,angle=-90]{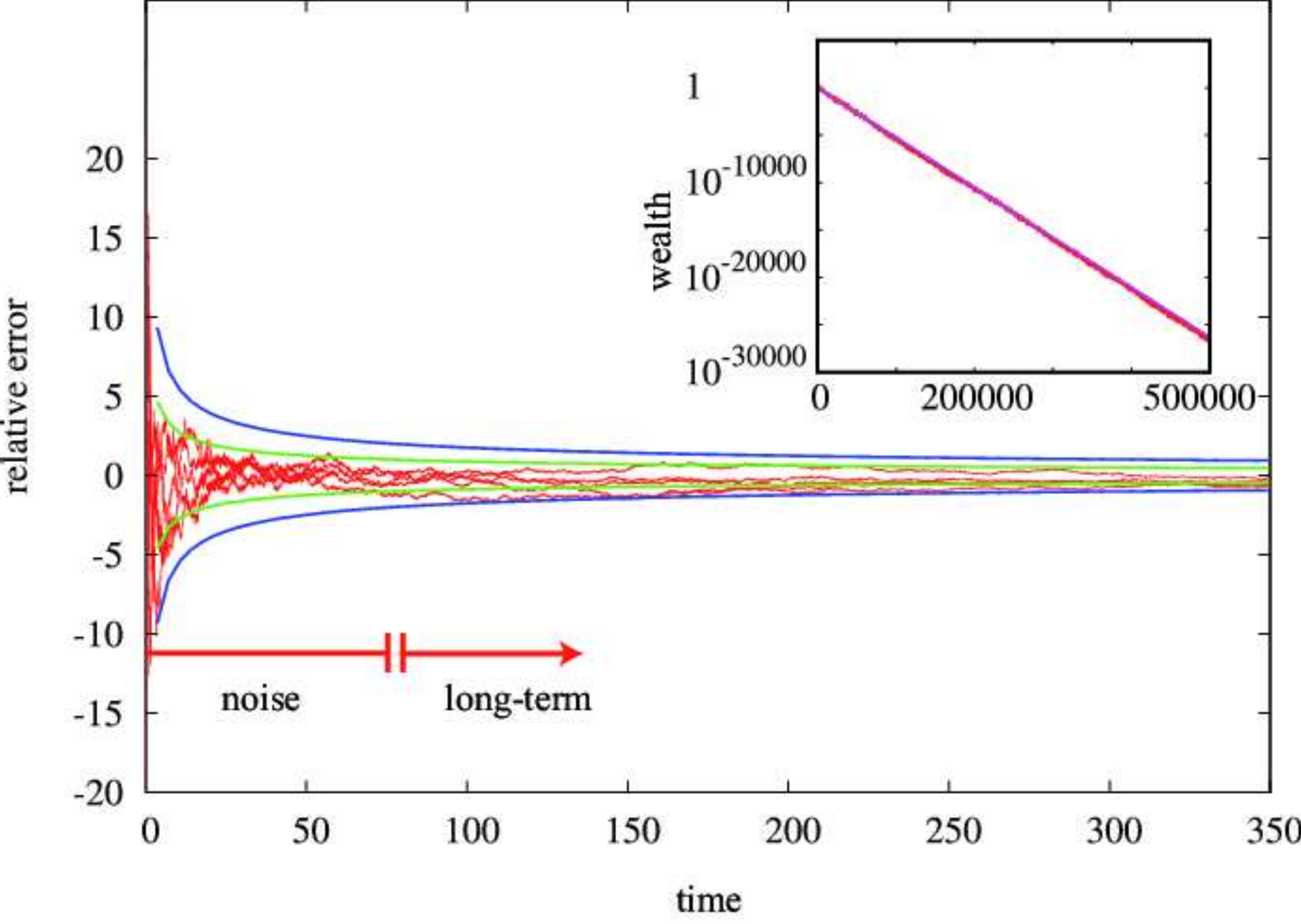}
\end{center}
\caption{ Relative errors
  $\frac{g_{\text{est}}(T,N=1)-\bar{g}}{\bar{g}}$ in estimates of the
  time-average growth rate, using the estimator $g_{\text{est}}(T,N=1)
  = \frac{1}{T} \ln \left(\frac{p(T)}{p(0)}\right)$ in single
  realizations of the process described in the caption of
  \fref{universes}. Green lines show one relative standard deviation
  from expected estimates, $\frac{\sigma T^{-1/2}}{\bar{g}}$, blue
  lines show two standard deviations.\newline \underline{Inset:}
  Long-time averages approach deterministic behavior with the
  time-average growth rate.}
\flabel{relative_error}
\end{figure}

\Eref{solution} shows that the median of
$\left(\frac{p(T)}{p(0)}\right)$ is
$\exp\left(\left(\mu-\frac{\sigma^2}{2}\right)T\right)$. But this is
not the expectation value, as the multiplicative nature of the process
makes for very large, though unlikely, positive fluctuations, and in
the ensemble {\it but not in time} these off-set the term
$-\frac{\sigma^2}{2}T$. Interpreting this in an investment context
exposes the dangers of misinterpreting \eref{return} and \eref{Kelly}:
using the ensemble-average growth rate where the time-average growth
rate would be appropriate overestimates the effect of positive
fluctuations. Extreme situations can be envisaged, where the investor
is bound to lose everything, although from the perspective of the
ensemble average, a few lucky copies of him in parallel universes make
up for his loss, making an investment proposition seem attractive
\citep{Peters2009}. But because resources cannot be exchanged with
other members of the ensemble (in parallel universes), this offset is
of no use to the investor as he progresses through time.

Using the estimator \eref{growth_estimate}, we have shown that
\eref{return} is an ensemble-average growth rate where stochasticity
is removed by the limit $N\to \infty$ and the effects of time are
suppressed. \Eref{Kelly} is the time-average growth rate, where
stochasticity is removed by the limit $T\to \infty$. The difference
between the two explicitly calculated growth rates, {\it i.e.} the
fact that the limits $\lim_{N\to\infty}$ and $\lim_{T\to\infty}$ do
not commute, is a manifestation of the non-ergodicity of the
system. Both rates can be obtained as ensemble-averages as in
\eref{expected} and \eref{growth_rate}; in differential form, the
ensemble-average growth rate is straight-forward, whereas the
time-average growth rate requires the application of \Ito's
formula. Intuitively, \Ito's formula corresponds to the inclusion of
effects of ignorance regarding the future (see \cite{LauLubensky2007},
\cite{Oksendal2005}, Ch.~3), and it encodes the effect of time
correctly for noise terms of the type of \eref{finite}. 

At zero volatility the ``time-average'' (certain) growth rate equals
the ensemble-average growth rate, and
any investor will choose the highest-yielding portfolio available,
maximizing both the time-average growth rate and the ensemble-average
growth rate over all possible universes (there is only one possible
universe now). Generalizing to the stochastic case, it is still
sensible to maximize time-average growth rates, but this is not
equivalent to maximizing ensemble-average growth rates. A good guide,
whether investing with or without volatility, is concern for the
future, rather than concern for copies of oneself in parallel
universes.

\section{Minimum investment horizon and optimal leverage}
\label{Time_scales}
Logarithmic utility, the Kelly criterion and time-average growth rates
are often associated with ``long-term investment''. The long term here
means a time scale that is long enough for the deterministic part of
the exponent in \eref{solution} to dominate over the noise. Applying
this terminology to the present case, one is investing either for the
long term, or in a regime where randomness dominates -- the latter
case may be described as ``gambling''. If in \eref{solution}, the
$W(T)$ is replaced by $\sqrt{\ave{W(T)^2}}=\sqrt{T}$, we estimate that
gambling stops, and the long term begins when
\begin{equation}
t>t_c=\frac{\sigma^2}{\left(\mu-\frac{\sigma^2}{2}\right)^2}.
\elabel{long_term}
\end{equation}
In \fref{universes} the corresponding time scale is $t_c \approx 75$
time steps, indicated by the break in the red arrow. At this
point, the typical relative error in estimates
$g_\text{est}(T,N=1)$ of $\bar{g}$ based on past performance is
unity, indicated by a break in the arrow in \fref{relative_error}. In
a single universe, such as our reality, the system is never dominated
by the ensemble-average growth rate -- neither in the short run nor in
the long run. Instead, there is an initial noise-regime where no
significant trends can be discerned, whereafter the time-average
growth rate dominates the performance.

\Eref{long_term} indicates how long we must expect to wait for the
trend of the market to become distinguishable from fluctuations,
wherefore $t_c$ may be viewed as a minimum investment horizon.  The
parameters of \fref{surface}, with a time unit of one year, imply
$t_c\approx 4.6$ years. Historical comparisons between portfolios with
similar stochastic properties are meaningful only on much longer time
scales.

The use of the null model of maximizing the time-average growth rate
eliminates the {\it a priori} need to specify
risk-preferences. Tailoring real-life investments to real investors'
needs does require difficult to formalize knowledge of their
circumstances, but a number of issues can be illuminated without such
knowledge in the simple context of the null-hypothesis. For instance,
a well-defined optimal leverage can be computed as will be shown
now. The result follows directly from Kelly's (1956) arguments, and
several authors have come to the same conclusions using different
methods \citep{Kestner2003,Thorp2006}. In addition, the
characteristic time scale of \eref{long_term} is calculated for the
leveraged case, which defines a critical leverage where the expected
growth rate vanishes.

Any efficient portfolio along the straight efficient frontier can be
specified by its fractional holdings of the market portfolio
\citep{Sharpe1964}, which we define as the leverage, $l$. For instance,
an investor who keeps all his money in the riskless asset holds a
portfolio of leverage $l=0$; half the money in the riskless asset and
half in the market portfolio is leverage $l=0.5$, and borrowing as
much money as one owns and investing everything in the market
portfolio corresponds to $l=2$ {\it etc}.

The ensemble-average growth rate in the leveraged case can be written
as the sum $\mu_{\text{riskless}}+ l \mu_{\text{excess}}$, where
$\mu_{\text{excess}}$ is the excess ensemble-average growth rate of
the market portfolio over the riskless growth rate, see
\fref{surface}. At zero leverage, only the riskless growth rate
enters; the excess ensemble-average growth rate is added in proportion
to the leverage. Noting that both the ensemble-average growth rate and
the volatility depend linearly on the leverage, we obtain the
leveraged stochastic process
\begin{equation}
dp_l(t)=p_l(t) \left(\mu_{\text{riskless}}+l \mu_{\text{excess}}) dt+ l \sigma_M dW_t\right),
\elabel{leverage_motion}
\end{equation}
where $\sigma_M$ is the volatility of the market portfolio. Just like
with \eref{motion}, we can use It\^{o}'s formula, \eref{formula}, to
derive the equation of motion for the logarithm of the price,
$p_l(t)$, of the leveraged portfolio,
\begin{equation}
d \ln p_l =\left(\mu_{\text{riskless}}+l\mu_{\text{excess}}-\frac{l^2\sigma_M^2}{2}\right)dt+l\sigma_M dW_t.
\end{equation}
The time-average leveraged exponential growth rate is thus\footnote{This
can also be seen immediately by replacing in \eref{growth_rate},
$\mu \to \mu_{\text{riskless}}+l\mu_{\text{excess}}$, $\sigma \to l\sigma_M$}
\begin{align}
\bar{g_l}&= \frac{1}{dt}\ave{ d \ln p_l}\elabel{l_growth_rate} \\
&=\left(\mu_{\text{riskless}}+l\mu_{\text{excess}}-\frac{l^2\sigma_M^2}{2}\right).\nonumber
\end{align}
The positive contribution to $\bar{g_l}$ is linear in the leverage,
but the negative contribution is quadratic in the leverage. The
quadratic term is an effect of time and makes the time-average growth
rate non-monotonic in the leverage.

Markowitz (1952) rejected strategies of maximum ensemble-average
growth because the corresponding portfolios are likely to be
under-diversified and hence to have an unacceptably high
volatility. \Eref{l_growth_rate} shows that in the limit of large
leverage, seeking high ensemble-average growth rates, the time-average
growth rate along the efficient frontier diverges negatively, as
$\lim_{l \to \infty} \bar{g_l}/l^2 =-\sigma_M^2/2<0$.

To find the optimal leverage, we differentiate
\eref{l_growth_rate} with respect to $l$ and set the result to
zero, obtaining
\begin{equation}
l_{\text{opt}}=\mu_{\text{excess}}/\sigma_M^2. 
\elabel{optimal_leverage}
\end{equation}
The second derivative of \eref{l_growth_rate} with respect to $l$ is
$-\sigma_M^2$, which is always negative, implying that $\bar{g_l}$
corresponding to $l_{\text{opt}}$ is maximized. This calculation shows
that there exists a privileged portfolio along the efficient
frontier. If the market portfolio has a lower volatility than the
portfolio of maximum time-average growth rate, as in \fref{surface},
then the wise investor will leverage his position by borrowing
($l_{\text{opt}}>1$). If, on the other hand, the market portfolio has
a higher volatility, as in \fref{universes}, then he will keep some
fraction of his money safe ($l_{\text{opt}}<1$).

We note that in geometric Brownian motion $\bar{g}$ can never be
increased by decreasing $\mu$ at constant volatility, nor can it be
increased by increasing volatility at constant $\mu$, because from
\eref{growth_rate}, $\frac{\partial \bar{g}}{\partial \mu}>0$ and
$\frac{\partial \bar{g}}{\partial \sigma}<0$. Therefore, the globally
({\it i.e.}  selected from all possible portfolios) growth-optimal
portfolio will be located on the efficient frontier. This temporal
optimization thus does not contradict modern portfolio theory. For the
dynamics of \eref{motion} it confirms \eref{efficient1} and
\eref{efficient2} as the definition of efficient, {\it i.e.}
potentially optimal, portfolios.

Including the leverage in \eref{long_term} results in the leveraged
characteristic time scale separating gambling from investing
\begin{equation}
t_c^l=\frac{l^2\sigma_M^2}{\left(\mu_{\text{riskless}}+l\mu_{\text{excess}}-\frac{l^2\sigma_M^2}{2}\right)^2}.
\elabel{long_term_l}
\end{equation}
This time scale diverges at the critical leverages,
\begin{equation}
l_c^{\pm}=l_{\text{opt}}\pm\sqrt{l_{\text{opt}}^2+2\left(\frac{\mu_{\text{riskless}}}{\sigma_M^2}\right)}
\elabel{l_crit}
\end{equation}
We are interested especially in the positive root\footnote{The
  negative root corresponds to zero expected growth rate in a
  negatively leveraged portfolio, consisting of the riskless asset and
  a small short position in the market portfolio, with the parameters
  in \fref{surface} this happens at $l_c^-\approx -0.80$.}, $l_c^+$,
where the time-average leveraged growth rate (the denominator of
\eref{long_term_l}) is zero due to over-leveraging. The minimum
investment horizon \eref{long_term_l} becomes infinite, meaning that
such an investment will forever be a gamble. The parameters of
\fref{surface}, for example, imply $l_c^+\approx 3.88$, where the
extrapolations of the white and green lines in the figure cross.

For leverages, $l>l_c^+$, the time-average growth rate is negative,
and the time scale \eref{long_term_l} is finite and marks the
transition between noise and discernible loss of invested capital.
This is the case in \fref{universes}, where $l_{\text{opt}}\approx
0.25$, the critical leverage $l_c\approx 0.49$, the system runs at
$l=1$, and $t_c^{l=1}\approx 75$ time units.

\subsection{Sharpe Ratio}
\label{Sharpe_Ratio}

The results from Sec.~\ref{Time_scales} also yield insights into the
Sharpe ratio, mentioned in Sec.~\ref{Introduction}. Writing in 1966,
Sharpe suggested to assess the quality of a portfolio specified by
some $\mu$ and $\sigma$ using the slope of the straight line in the
$\mu$ {\it vs.} $\sigma$ plane in \fref{surface} that corresponds to
combinations of the portfolio and the riskless asset ({\it
  i.e.} to all possible values of $l$),
\begin{equation}
  S=\frac{\mu-\mu_{\text{riskless}}}{\sigma}.  \elabel{Sharpe}
\end{equation}
The concept of ergodicity is relatively young, with major results in
ergodic theory emerging in the second half of the 20$^{\text{th}}$
century \citep{LebowitzPenrose1973}. For early roots of the discussion,
see \cite{Uffink2004}. The concepts did not immediately diffuse into
the economics literature: in the 1966 paper Sharpe makes no
distinction between time- and ensemble-averages. It is assumed here
that he refers to ensemble averages throughout the paper. Some authors
assume that he refers to time-averages \citep{BouchaudPotters2000}, but
the resulting quantity,
$\frac{\mu-\frac{\sigma^2}{2}-\mu_{\text{riskless}}}{\sigma}$, has a less
straight-forward meaning. Despite the exclusive use of ensemble
averages in \eref{Sharpe}, $S$ is also meaningful in the context of
time averages in geometric Brownian motion: given two portfolios $M_1$
and $M_2$, where $S(M_1)>S(M_2)$, the optimally leveraged portfolio
$M_1^{l_{\text{opt 1}}}$ always has a greater time-average growth rate
than the optimally leveraged portfolio $M_2^{l_{\text{opt 2}}}$.

Sharpe was fully aware of the limitations of his measure:
$S(M_1)>S(M_2)$ does not mean that an investment in $M_1$ will
outperform an investment in $M_2$, as it is possible that $M_1$ is far
from optimally leveraged. He concluded that ``The investor's task is
to select from among the efficient portfolios the one that he
considers most desirable [{\it i.e.} to choose a leverage $l$], based
on his particular feelings regarding risk and expected return
\citep{Sharpe1966}.'' Without considerations of ergodicity, investors
are indeed left to making decisions based on their feelings.

The optimal leverage
$l_{\text{opt}}=\frac{\mu_{\text{excess}}}{\sigma_M^2}$, differs from
the Sharpe ratio \eref{Sharpe} for the market portfolio only by a
square in the volatility. Indeed, it may also be considered a
fundamental measure of the quality of a portfolio: if the optimal
leverage for a given investment opportunity is high, then this is a
good opportunity that calls for a large commitment.

Optimal leverage, unlike the Sharpe ratio, is a dimensionless
quantity. This is a significant difference, as it implies that the
numerical value of optimal leverage, which is a pure number, can
distinguish between fundamentally different dynamical regimes, see
\eg  \cite{Barenblatt2003}. For example, a value
$l_{\text{opt}}<1$, irrespective of its constituting $\sigma_M$ and
$\mu_{\text{excess}}$, or the unit of time used to measure these
quantities, implies that an investor will be better off keeping some
of his money safe. The Sharpe ratio, \eref{Sharpe}, on the other hand,
has dimension $[S]=\mathrm{T}^{-1/2}$, wherefore it depends on the chosen unit
of time, implying that its numerical value is arbitrary. For example,
a portfolio with Sharpe ratio 5, where $\mu$ is measured as a percentage
per year and $\sigma$ as a percentage per square-root of one year
would have Sharpe ratio $\frac{5}{\sqrt{365}}\approx 0.26$ if the
chosen time unit were one day.

In the context of the current crisis the limitation of the Sharpe
ratio is perhaps best expressed by its insensitivity to leverage,
\begin{equation}
S_l=\frac{l(\mu-\mu_{\text{riskless}})}{l\sigma}=S.
\elabel{Sharpe_l}
\end{equation}
Using the Sharpe ratio alone to assess the quality of an investment
would be dangerous, as this measure cannot detect the negative effects
of leverage. Given the systemic incentives for using large leverage,
this insensitivity can become a danger to market stability.

\section{Discussion}
\label{Discussion}
Practically relevant lessons from the above considerations may be
learned from the extremes. A 100\% mortgage, for instance, corresponds
to infinite leverage, implying $\bar{g} \to -\infty$, on the
borrower's investment (assuming that the purchase is not part of a
larger investment portfolio). Although total loss on a home purchase
only means dipping into negative equity, certain financial products
that have become popular in recent years must be regarded as
irresponsible. Conversely, \eref{l_growth_rate} shows that for $l\ll
1$ the time-average growth rate is well approximated by the ensemble
average. From the very beginning, treatments of gambling focused on
ensemble-average outcomes, \eg Cardano's 16$^{\text{th}}$ century
``Liber de Ludo Aleae'', translated by Gould in \cite{Ore1953}. This
is appropriate as long as wagers are much less than total wealth,
meaning leverage close to zero, where $\bar{g}$  becomes
indistinguishable from $\ave{g}$.

The current financial crisis started in the summer of 2007, with the
US housing market collapsing and the visible consequence of Northern
Rock in the UK suddenly unable to raise credit, {\it i.e.} leverage,
on the open market. Subsequently, credit markets began to
freeze. After the nationalization of Fannie Mae and Freddie Mac and
the bankruptcy of Lehman Brothers in September 2008 entire markets for
leverage-oriented financial products disappeared (securitization and
credit default swaps, for instance, were strongly affected). Leverage
clearly played a big r\^ole. The scale of this crisis suggests to
revisit some of the basic tenets of the economic formalism, including
the concept of equilibrium, the r\^ole of time, and indeed the
frequent implicit assumption of ergodicity.

It is emphasized here that ergodicity can be inadvertently assumed by
writing an expectation value $\ave{}$, which implies the limit
$N\to\infty$ of infinitely many realizations of a process. The
straight-forward expected outcome, $\ave{p(T)}=\int p P(p) dp$, of
some investment is indeed an average over many universes. Even if
subsequently an exponential growth rate is derived from this as
$\frac{1}{T}\ln\left(\frac{\ave{p(T)}}{p(0)}\right)$, the
multiplicative effects of time are ignored, and the result will not be
the time-average growth rate. We have seen that this problem persists,
even if \Ito's formula is used to find the distribution of $p(T)$.

Real portfolios of constant leverage (apart from $l=0$ and possibly
$l=1$) need to be constantly rebalanced as the value of the market
portfolio fluctuates and changes the fraction of wealth invested in
it. Holding any such portfolio is costly, both in terms of monitoring
time and in terms of transaction costs. For applications, the above
considerations would thus need to be adapted, even if real prices were
perfectly described by \eref{motion}. The value of real optimal
leverage depends on the investor's ability to balance portfolios,
which is affected by the available technology and, due to market
impact, by the volume of the investment. The assumptions made in this
study are likely to lead to an over-estimate of optimal leverage:
\Eref{optimal_leverage} was derived in continuous time, corresponding
to truly constantly rebalanced portfolios, zero transaction costs were
assumed, log-normal return-distributions, certain knowledge of $\mu$ and
$\sigma$, and no risk premiums charged on money borrowed for
leveraging.

Modelling the S\&P500 or the DJIA with \eref{motion}, one would choose
parameters close to those of \fref{surface} with time units of one
year. This implies an optimal leverage, as calculated above, of 1.54,
but it is unlikely that the simple strategy of borrowing money and
investing it in the S\&P500 would outperform the market. It is equally
unlikely that investing only part of one's money in the S\&P500 would
outperform the market, as would be the case if $l_{\text{opt}}<1$. A
reasonable guess is that real optimal leverage is close to
$l_{\text{opt}}=1$, a possible attractor for a self-organized market
system. How could such statistical market efficiency work? If
$l_{\text{opt}}>1$, money will be borrowed to be invested. This
situation can arise as a consequence of low interest rates and
low-cost credit. Leverage tends to increase volatility due to
potential margin calls and similar constraints on investors
\citep{Geanakoplos1997}. Thus, as investors increase their leverage,
they reduce optimal leverage, creating a negative feedback loop whose
strength depends on the magnitude of the impact of leverage on
volatility (``$\frac{d\sigma}{dl}$''), this brings optimal leverage
down, $l_{\text{opt}} \to 1^+$. Conversely, if optimal leverage is
less than unity, investors will sell risky assets, thereby reducing
prices and increasing expected returns, such that optimal leverage
increases, possibly up to $l_{\text{opt}} \to 1^-$.

The time scales associated with such a feedback loop can be long,
especially in situations where leverage initially reduces
volatility. The current financial crisis has been related to a
continued extension of credit \citep{Soros2008c,Soros2008}, {\it i.e.}
increasingly leveraged investments.  Effects of leverage that are
  initially volatility-reducing can be discussed in terms of
  mortgages: easy availability of mortgages increases house prices,
  which leads to few defaults, even if loans are given to borrowers
  who cannot service them from wages. In turn, because volatility
  decreases, optimal leverage increases, and houses appear a good
  investment. Consequently, more money is lent to home buyers, leading
  to a destabilizing run-away dynamics. Soros (2008a) has called such
  interaction between the asset price and the investment in the asset
  (the loan) ``reflexivity''. After this initial reflexive phase of
  bubble creation, leverage will be perceived in some areas to have
  risen far beyond optimality, creating an unstable market situation.
The ensuing crisis may be viewed as a response where volatility
suddenly increases, reducing optimal leverage, which in turn leads to
deleveraging and falling prices.

The use of leverage is not fundamentally constrained by the prevailing
framework of portfolio selection, which relies on a necessarily and
explicitly subjective notion of optimality, dependent on utility, or
risk preferences. This has become problematic because asymmetric
reward structures have encouraged excessive
leveraging. Securitization, for example, creates such structures by
separating the sellers of leverage products (such as mortgages), who
are rewarded for every sale, from those who eventually bear the
risk. Similarly, an investment manager who benefits from gains in the
account he manages but is not personally liable for losses has an
incentive to exceed growth-optimal leverage (see footnote on
p.~\pageref{horse_note}). In the ideal setup discussed above, the
growth-optimal ansatz suggests a simple reward scheme through
alignment of interests: requiring investment managers to invest all
their wealth in the accounts managed by them. It is thereby achieved
that the growth-optimal investment strategy for the account is also
growth optimal for the investment manager. It is commonly said that
excessive leverage arises when investors are short-term oriented, but
there is no benefit from leveraging beyond optimality even in the
short term -- this regime is dominated by noise, not by
ensemble-average growth rates. It is harmless to reward investment
managers daily, hourly or indeed continuously for their performance --
as long as they share the risks as much as the rewards.

\Eref{l_growth_rate} carries an important message regarding reward
structures in the financial industry. Excessive leverage leads to
large fluctuations in asset prices but also more generally in economic
output (the current recession being a case in point). The introduction
of such fluctuations must reduce time-average economic
growth. Remuneration practices have been criticized on a moral basis,
using concepts like greed, excess and inequality. Objectively one can
argue that remuneration structures can go against the common good by
reducing economic growth by generating unnecessary fluctuations.

In conclusion, utility functions were introduced in the early
18$^{\text{th}}$ century to solve a problem that arose from using
ensemble averages where time averages seem more appropriate. Much of
the subsequently developed economic formalism is limited by a similar
use of ensemble averages and often overlooks the general problem that
time- and ensemble averages need not be identical. This issue was
treated in detail only in the 20$^{\text{th}}$ century in the field of
ergodic theory. Making use of this work, a privileged portfolio
uniquely specified by an optimal leverage and a maximized time-average
growth rate is seen to exist along the efficient frontier, the
advantages of which have also been discussed elsewhere
\citep{Breiman1961,MertonSamuelson1974,CoverThomas1991}. The concept of
many universes is a useful tool to understand the limited significance
of ensemble averages. While modern portfolio theory does not preclude
the use of, in its nomenclature, logarithmic utility, it seems to
underemphasize its fundamental significance. It was pointed out here
that the default choice to optimize the time-average growth rate is
physically motivated by the passage of time and the non-ergodic nature
of the multiplicative process.

\section*{Acknowledgements}
I thank B. Meister for introducing me to Kelly's work and for
hospitality at Renmin University, Beijing, D. Farmer for many
insightful discussions and hospitality at the Santa Fe Institute (SFI)
under the Risk Program, Y. Schwarzkopf for making me aware of
reference~\citep{Timmermann1993}, F. Cooper for useful comments after
a presentation at SFI, B. Hoskins for a careful reading of the
manuscript, G. Pavliotis for discussions regarding stochastic
processes, J. Anderson and C. Collins of Baillie-Gifford~\& Co.,
A. Rodriguez, M. Mauboussin, and W. Miller of Legg Mason Capital
Management and A. Adamou for insightful discussions, and G. Soros for
pointing me to references \citep{Soros2008c,Soros2008}. This work was
supported by Baillie-Gifford~\& Co.  and ZONlab Ltd.; the Risk Program
at SFI is funded by a generous donation from W. Miller.
\bibliographystyle{rQUF}
\bibliography{/Users/obp48/bibliography/bibliography}
\end{document}